\documentstyle[12pt]{article}
\begin{document}

\title{Can neutrino vacuum support the wormhole?}
\author{V. Khatsymovsky \\
 {\em Budker Institute of Nuclear Physics} \\ {\em Novosibirsk,
 630090,
 Russia} \\ {\em E-mail address: khatsym@inp.nsk.su}}
\date{}
\maketitle
\begin{abstract}
The renormalised vacuum expectation values of massless fermion
(conventionally call it neutrino) stress-energy tensor are
calculated in the static spherically-symmetrical wormhole
topology. Consider the case when the derivatives of metric
tensor over radial distance are sufficiently small (in the scale
of radius) to justify studying a few first orders of
quasiclassical (WKB) expansion over derivatives. Then we find
that violation of the averaged weak energy condition takes place
irrespectively of the detailed form of metric. This is a
necessary condition for the neutrino vacuum to be able to
support the wormhole geometry. In this respect, neutrino vacuum
behaves like electromagnetic one and differs from the conformal
scalar vacuum which does not seem to violate energy conditions
for slowly varied metric {\em automatically}, but requires
self-consistent wormhole solution for this.
\end{abstract}
\newpage
{\bf 1.Introduction.} The possibility of existence of static
spherically-symmetrical \\ traversible wormhole as
topology-nontrivial solution to the Einstein equations has been
first studied by Morris and Thorne in 1988 \cite{MT}. They have
found that the material which threads the wormhole should violate
weak energy condition at the throat of the wormhole, that is,
radial pressure should exceed the density. Moreover, Morris,
Thorne and Yurtsever \cite{MTY} have pointed out that averaged
weak energy condition (i.e. that integrated over the radial
direction) should also be violated. Since that time much
activity has been developed in studying the wormhole subject
(see, e.g., review by Visser \cite{Vis}) of which we consider
here the possibility of existence of self-consistent wormhole
solutions to semiclassical Einstein equations. Checking this
possibility requires finding vacuum expectation value of the
stress-energy tensor as functional of geometry and solving the
Einstein equations with this tensor as a source.

Recently self-consistent spherically symmetric wormhole
solution has been found numerically by Hochberg, Popov and
Sushkov \cite{HPS} for the quantised scalar field vacuum playing
the role of a source for the gravitation. These authors employ
vacuum expectation values of the stress-energy tensor for the
scalar field found by Anderson, Hiscock and Samuel \cite{AHS}
although Anderson, Hiscock and Taylor have argued (without
solving the backreaction problem, however) that these values for
massive minimally and/or conformally coupled scalars fail to
meet the requirements for maintaining five particular types of
static spherically symmetric wormholes \cite{AHT}.
As for the experimentally known fields, we have calculated in
\cite{Kh} the renormalised stress-energy tensor of
electromagnetic vacuum with the help of the covariant geodesic
point separation method of regularisation \cite{Christ}. It has
been found to violate weak energy condition at the wormhole
throat in the first nonvanishing order in the expansion over the
derivatives of metric (the WKB expansion). This is a necessary
condition of existence of self-maintained wormhole solution.
Important is that this violation takes place irrespectively of
the detailed form of metric.

Also violation of the averaged weak energy condition is
necessary (but not sufficient, of course) condition for the
existence of vacuum self-maintained wormhole. The validity of
the weak averaged energy condition for the self-consistent
solution to the Einstein equations has been studied by Flanagan
and Wald \cite{FW} for the massless scalar field. Flanagan and
Wald have found that averaged weak energy condition holds for
self-consistent solution if being additionally averaged
transverse to the geodesic using a suitable Plank scale smearing
function. The Flanagan-Wald result is obtained in the context of
perturbation theory about flat spacetime, i.e. for the Minkowski
topology. In our preceding paper \cite{Kh1} dealing with
electromagnetic field we note that validity of the averaged weak
energy condition is substantially defined by the space(-time)
topology. This condition is violated in the first nonvanishing
order of WKB expansion over derivatives of metric (over radial
coordinate) if topology is that of wormhole.

Important is possibility to have macroscopic wormhole size under
proper conditions. Flanagan and Wald \cite{FW} and Ford and
Roman \cite{FR} have argued that wormhole size would be
Plank-scale. These arguments are based on the assumption that
coefficient at the curvature squared (Weyl term) in the
effective gravity action is of Plank scale value. Meanwhile,
experimental bounds on this coefficient are not very restrictive
\cite{DeWitt}, and in \cite{Kh1} we speculate that this allows
the wormhole size to be as large as radius of the Sun. A
possible large value of the Weyl term might be provided by
infra-red contribution into effective action from the massless
fields, such as electromagnetic one.

In the given note we perform analogous calculations for the
stress-energy of the neutrino vacuum in the wormhole topology.
We find that violation of the energy conditions takes place in
the self-consistent manner analogously to what occurs in the
electromagnetic vacuum in the wormhole topology \cite{Kh, Kh1}.
This is the necessary condition and justifies further work
towards construction of self-consistent wormhole solution in the
real physical vacuum of spin 1/2 and 1 fields.

\bigskip
{\bf 2.Calculation.} The notations we employ are mainly those of
ref.\cite{DeWitt1} used in ref.\cite{Christ}. The notations for
the metric functions $r(\rho)$, $\Phi(\rho)$ are read from the
following expression for the line element:

\begin{equation}
ds^2=r^2(\rho)(d\theta^2+\sin^2{\!\theta}\,d\phi^2)+d\rho^2
-\exp{(2\Phi)}dt^2.
\end{equation}
The $\mu,\nu,\lambda,...$ denote world tensor indices $t,\rho,
\theta,\phi$ while $\alpha,\beta,\gamma,...$ are purely space
such indices. We also need notation for the local flat space
indices $a,b,c,...=1,2,3$. In the case of spherical symmetry it
is convenient to use decomposition of spinor field over
spherical waves. Therefore we consider neutrino field as
particular vanishing-mass case of the four-component Dirac field
and take $\gamma$-matrices in the local flat Minkowski
coordinates $x^0,x^1,x^2,x^3$ in the usual standard
representation:

\begin{equation}
\gamma_0=\left (\matrix{i~~~0\cr 0 -i\cr}\right ),~~~
\gamma_a=\left (\matrix{0~~~~~i\sigma_a\cr -i\sigma_a~~0\cr}
\right ),
\end{equation}
$\sigma_a$ being Pauli matrices, $\sigma_1\sigma_2=i\sigma_3$,
\dots. We choose tetrad by requiring that $\gamma_{\mu}$ were
proportional to the expressions which would follow if
$x^0,x^1,x^2,x^3$ were treated as usual Cartesian coordinates
$t,x,y,z$ and $t,\rho,\theta,\phi$ were treated as usual
spherical flat coordinates:

\begin{eqnarray}
&&\gamma_t=\gamma_0\exp{(\Phi)},\nonumber\\
&&\gamma_{\rho}=\gamma_1\sin{\theta}\cos{\phi}+\gamma_2
\sin{\theta}\sin{\phi}+\gamma_3\cos{\phi},\nonumber\\
&&\gamma_{\theta}=(\gamma_1\cos{\theta}\cos{\phi}+\gamma_2
\cos{\theta}\sin{\phi}-\gamma_3\sin{\phi})r,\\
&&\gamma_{\phi}=(-\gamma_1\sin{\phi}+\gamma_2\cos{\phi})r
\sin{\theta}.\nonumber
\end{eqnarray}
The $\sigma_{\alpha}$ are defined by analogous formulas with
$\gamma$ replaced by $\sigma$. It proves convenient to write
bispinor spherical waves in the form

\begin{eqnarray}
&\psi={\exp{(-\Phi/2)}\over r}\left (\matrix{\eta_+\Omega_+\cr
\eta_-\Omega_-\cr}\right )\exp{(-i\omega t)},&\nonumber\\
&\Omega_{-lm}=
\left (\matrix{-\sqrt{l-m\over 2l+1}Y_{lm}\cr \sqrt{l+m+1\over
2l+1}Y_{l,m+1}}\right ),~~\Omega_{+lm}=
\left (\matrix{\sqrt{l+m\over 2l-1}Y_{l-1,m}\cr \sqrt{l-m-1\over
2l-1}Y_{l-1,m+1}}\right )&
\end{eqnarray}
and analogous one with $\eta_-\Omega_-$ and $\eta_+\Omega_+$
interchanged. It is implied that spherical function $Y_{lm}=0$
at $m>l$. Here $l\geq 1$ and the full momentum $j=l-1/2$ with
$m+1/2$ being it's projection. The Dirac conjugate is
$\overline{\psi}=\psi^+\gamma^0$ and spherical spinors satisfy

\begin{equation}
\sigma_{\rho}\Omega_{\pm}=\pm i\Omega_{\mp},~~(\sigma^{\theta}
\partial_{\theta}+\sigma^{\phi}\partial_{\phi})\Omega_{\pm}=-i
{l\mp 1\over r}\Omega_{\mp},~~\sum_m{\Omega^+_{\pm}\Omega_{\pm}}
={l\over 2\pi}.
\end{equation}
The covariant derivative of a spinor $\psi_{;\mu}\equiv
D_{\mu}\psi=\psi_{,\mu}-\Gamma_{\mu}\psi$ can be defined from
the requirement

\begin{equation}
\gamma_{\mu;\nu}\equiv\gamma_{\mu,\nu}-\Gamma^{\lambda}_{\mu\nu}
\gamma_{\lambda}-\Gamma_{\nu}\gamma_{\mu}
+\gamma_{\mu}\Gamma_{\nu}=0.
\end{equation}
This gives

\begin{equation}
D_t=\partial_t+{1\over 2}\Phi^{\prime}\gamma_t\gamma_{\rho},~~
D_{\rho}=\partial_{\rho},~~D_A=\partial_A
+{r^{\prime}-1\over 2r}\gamma_A\gamma_{\rho},~~A=\theta,\phi.
\end{equation}
Description of the formalism is completed by setting the
standard expression for the action

\begin{equation}
S={i\over 4}\int{(\overline{\psi}\gamma^{\mu}\psi_{;\mu}
-\overline{\psi_{;\mu}}\gamma^{\mu}\psi)\sqrt{-g}d^4x}
\end{equation}
which provides the stress-energy

\begin{equation}
T_{\mu\nu}=-{i\over 4}(\overline{\psi}\gamma_{\mu}\psi_{;\nu}
+\overline{\psi}\gamma_{\nu}\psi_{;\mu}).
\end{equation}
The equation of motion $\gamma^{\mu}\psi_{;\mu}=0$ in terms of
the two-component function $\eta=\left (\matrix{\eta_+\cr
\eta_-\cr}\right )$ takes the form (and the same for $\psi$ with
$\eta_-\Omega_-$ and $\eta_+\Omega_+$ interchanged):

\begin{equation}
\label{eta-eq}
\omega\left (\matrix{\eta_+\cr\eta_-\cr}\right )={d\over dz}
\left (\matrix{\eta_-\cr -\eta_+\cr}\right )+lU\left (\matrix{
\eta_-\cr\eta_+\cr}\right )\equiv{\cal O}\left (\matrix{\eta_+\cr
\eta_-\cr}\right )
\end{equation}
where we have denoted $U=\exp{(\Phi)}/r$ and introduced the new
variable $z$ via $dz=\exp{(-\Phi)d\rho}$.

Symmetrical separation of the points $x,\tilde{x}$ according to
the Christensen's prescription \cite{Christ} for the regularised
form of the stress-energy tensor gives, e.g.,

\begin{equation}
\label{T-split}
T^{\rm reg}_{tt}=-{i\over 4}[\overline{\psi}\gamma_t
(\tilde{\psi_{;t}})_x+\overline{(\tilde{\psi})_x}\gamma_t
\psi_{;t}-\overline{\psi_{;t}}\gamma_t(\tilde{\psi})_x
-\overline{(\tilde{\psi_{;t}})_x}\gamma_t\psi].
\end{equation}
Here $(\tilde{\psi})_x$, $(\tilde{\psi_{;t}})_x$ are the fields
$\psi (\tilde{x})$, $\psi_{;t}(\tilde{x})$, respectively,
transported in parallel way from $\tilde{x}$ to $x$ along the
geodesic. We split the point in radial direction so that
$x=(t,\rho,\theta,\phi)$, $\tilde{x}=(t,\tilde{\rho},\theta,
\phi)$, $\tilde{\rho}-\rho=\epsilon\rightarrow 0$.

Now we substitute $\psi$ as the field operator expanded in terms
of creation and annihilation operators into (\ref{T-split}). As
in \cite{Christ}, the operator ordering $\psi^+\dots\psi
\Rightarrow{1\over 2}(\psi^+\dots\psi -\psi\dots\psi^+)$ is
implied. The vacuum expectation values of $T^{\rm reg}_{\mu\nu}$
will be denoted as components themselves: this will not lead to
any confusion. These values are

\begin{eqnarray}
\label{T-reg}
\left (\matrix{T^t_t\cr T^{\rho}_{\rho}\cr T^{\theta}_{\theta}
\cr}\right )^{\rm reg}={1\over 4\pi r\tilde{r}}\exp{\left (
-{\Phi\over 2}-{\tilde{\Phi}\over 2}\right )}\left\{\left (
\matrix{1\cr -1\cr 0\cr}\right )\left [\left (\exp{(-\Phi)}
+\exp{(-\tilde{\Phi})}\right )S_1\right.\right.\nonumber\\
\left.\left.-{\tilde{\Phi^{\prime}}
-\Phi^{\prime}\over 2}S_a\right ]+\left (\matrix{0\cr 1\cr 
-{1\over 2}\cr}\right )\left [\left ({1\over r}
+{1\over\tilde{r}}\right )S_s+\left ({\tilde{r^{\prime}}\over
\tilde{r}}-{r^{\prime}\over r}\right )S_a\right ]\right\}.
\end{eqnarray}
The tilde denotes the function at $\tilde{\rho}$. Here we
introduce the three basic sums over quantum numbers $n$ (which
labels the solutions to the eq.(\ref{eta-eq})) and $l$.(The
summation over $m$ is made in closed form thus eliminating the
angle $\theta,\phi$ dependence and leading to
$T^{\phi}_{\phi}=T^{\theta}_{\theta}$). These sums are

\begin{eqnarray}
\label{S}
S_1=\sum_{n,l}{l\omega\theta (\omega)(\eta_+\tilde{\eta}_+
+\eta_-\tilde{\eta}_-)},\nonumber\\
S_s=\sum_{n,l}{l^2\theta (\omega)(\eta_+\tilde{\eta}_-
+\eta_-\tilde{\eta}_+)},\\
S_a=\sum_{n,l}{l\theta (\omega)(\eta_+\tilde{\eta}_-
-\eta_-\tilde{\eta}_+)}.\nonumber
\end{eqnarray}
Here indices $n,l$ at $\eta_{\pm}$ and $\omega$ are suppressed,
$\theta (\omega)={1\over 2}\left (1+{\omega\over |\omega |}
\right )$, and $\eta$ is normalised as

\begin{equation}
\label{norm}
\int_{-\infty}^{+\infty}{(\eta_+^2+\eta_-^2)dz}=1
\end{equation}
(we use real functions $\eta_{\pm}$).

The operator ${\cal O}$ in (\ref{eta-eq}) is Hermitian w.r.t. the
internal product like (\ref{norm}) and can be made self-adjoint
by imposing appropriate boundary conditions at $z=\pm M$; then
we put $M\rightarrow\infty$. The two-component eigenvector
$\eta$ of ${\cal O}$ form the orthogonal set and by the
completeness property

\begin{equation}
\sum_n{\eta (z)\eta^+(\tilde{z})}={\bf 1}\cdot\delta
(z-\tilde{z})
\end{equation}
(here ${\bf 1}$ is unit $2\times 2$ matrix). Thereby summation
over $n$ in the stress-energy is reduced to finding the kernel
of the operator functions $f({\cal O})={\cal O}\theta ({\cal O})$
or $\theta ({\cal O})$:

\begin{equation}
\sum_n{f(\omega)\eta (z)\eta^+(\tilde{z})}
=f({\cal O})\delta(z-\tilde{z}).
\end{equation}
These functions can be rewritten as

\begin{equation}
\theta ({\cal O})={1\over 2}\left (1+{{\cal O}\over
\sqrt{{\cal O}^2}}\right ),~~{\cal O}\theta ({\cal O})
={1\over 2}({\cal O}+\sqrt{{\cal O}^2}).
\end{equation}
Nonlocality upon the action on $\delta (z-\tilde{z})$ arises
from the terms with $\sqrt{{\cal O}^2}$. Corresponding result at
$\tilde{z}\neq z$ can be expressed in terms of Green function
$G(s,l;z,\tilde{z})$ which satisfies

\begin{equation}
-G_{zz}^{\prime\prime}+\omega^2(z)G=\delta (z-\tilde{z}),
~~\omega^2(z)\equiv s^2+l^2U^2+lU_z^{\prime}\sigma_3
\end{equation}
in the following way:

\begin{eqnarray}
&&\sum_n{\eta (z)\eta^+(\tilde{z})\omega\theta (\omega)}
=-\int_0^{\infty}{{s^2ds\over\pi}G(s,l;z,\tilde{z})},\nonumber\\
&&\sum_n{\eta (z)\eta^+(\tilde{z})\theta (\omega)}
=\left (i\sigma_2{d\over dz}+lU\sigma_1\right )
\int_0^{\infty}{{ds\over\pi}G(s,l;z,\tilde{z})}.
\end{eqnarray}
Since the matrix $\omega^2(z)$ is diagonal, the WKB expansion
for $G$ over the powers of $\Delta z=\tilde{z}-z$ and over the
derivatives of $\omega^2(z)$ does not differ from that obtained
for $c$-number $\omega^2(z)$ in our previous paper \cite{Kh1}:

\begin{eqnarray}
2G(z,\tilde{z})=\exp{(-\omega\Delta z)}\left\{{1\over\omega}-
{1\over 8}{(\omega^2)^{\prime\prime}\over\omega^5}+{5\over 32}
{(\omega^2)^{\prime 2}\over\omega^7}\right.\nonumber\\
+\Delta z\left [-{1\over 4}{(\omega^2)^\prime\over\omega^3}-
{1\over 8}{(\omega^2)^{\prime\prime}\over\omega^4}+{5\over 32}
{(\omega^2)^{\prime 2}\over\omega^6}\right ]\nonumber\\
+\Delta z^2\left [-{1\over 4}{(\omega^2)^\prime\over\omega^2}-
{1\over 8}{(\omega^2)^{\prime\prime}\over\omega^3}+{5\over 32}
{(\omega^2)^{\prime 2}\over\omega^5}\right ]\nonumber\\
\left.+\Delta z^3\left [-{1\over 12}{(\omega^2)^{\prime\prime}
\over\omega^2}+{5\over 48}{(\omega^2)^{\prime 2}\over\omega^4}
\right ]+\Delta z^4{1\over 32}{(\omega^2)^{\prime 2}\over
\omega^3}\right\}.
\end{eqnarray}
Here the derivatives are over $z$. One need to additionally
expand over $U_z^{\prime}$ entering $\omega^2(z)$.

Substitute this into the expression for the stress-energy and
introduce instead of $s$ a new integration variable $q=\Delta
z\sqrt{s^2l^{-2}+U^2}$. Thereby we obtain a collection of terms
containing powers of derivatives of $U$ over $z$ (that is, of
$r$ and $\Phi$ over $\rho$) times coefficients of the type

\begin{equation}
\Delta z^k\int_{U\Delta z}^{\infty}{dq\over q^p\sqrt{q^2
-U^2\Delta z^2}}{d^j\over dq^j}{f(q)\over q^2}
\end{equation}
where

\begin{equation}
f(q)\equiv q^2\sum_{l=1}^{\infty}{l\exp{(-ql)}}
={q^2\over 4\sinh^2{q\over 2}}.
\end{equation}
Expansion of the sums (\ref{S}) in powers of $\Delta z$ is
achieved by carefully expanding $f(q)$ in Taylor series around
$q=0$.

When substituting $S_1,S_s,S_a$ into (\ref{T-reg}) one should
also expand $\tilde{r}$ and $\tilde{\Phi}$ over $\Delta z$ and
express $\Delta z$ in terms of $\epsilon =\Delta\rho$. The
result is (the derivatives are over $\rho$ now):

\begin{eqnarray}
&&\hspace{-15mm}8\pi^2r^4T_t^{t,{\rm reg}}
=-8{r^4\over\epsilon^4}
+{1\over 3}{r^2\over\epsilon^2}\left (1-r^{\prime 2}+2r^{\prime
\prime}r\right )-{1\over 3}{r\over\epsilon}r^{\prime}
+{1\over 60}\ln{{Lr\over\epsilon}}+{23\over 72}r^{\prime 2}
-{5\over 36}r^{\prime\prime}r,\\
&&\hspace{-15mm}8\pi^2r^4T_{\rho}^{\rho,{\rm reg}}=+24{r^4\over
\epsilon^4}-{1\over 3}{r^2\over\epsilon^2}\left [1+r^2\left (
2\Phi^{\prime\prime}+2\Phi^{\prime 2}-2\Phi^{\prime}{r^{\prime}
\over r}-{r^{\prime 2}\over r^2}+4{r^{\prime\prime}\over r}
\right )\right ]+{1\over 3}{r\over\epsilon}r^{\prime}\nonumber\\
&&\hspace{10mm}+{1\over 60}\left (\ln{{Lr\over\epsilon}}-1
\right )+{1\over 72}r^2\left (-2\Phi^{\prime\prime}
-2\Phi^{\prime 2}+2\Phi^{\prime}{r^{\prime}\over r}
-13{r^{\prime 2}\over r^2}\right ),\\
&&\hspace{-15mm}8\pi^2r^4T_{\theta}^{\theta,{\rm reg}}=-8{r^4
\over\epsilon^4}+{1\over 3}{r^2\over\epsilon^2}\left (
\Phi^{\prime\prime}+\Phi^{\prime 2}-\Phi^{\prime}{r^{\prime}
\over r}+{r^{\prime\prime}\over r}\right )-{1\over 60}\left (
\ln{{Lr\over\epsilon}}-{1\over 2}\right )\nonumber\\
&&\hspace{10mm}+{1\over 72}r^2\left (\Phi^{\prime\prime}
+\Phi^{\prime 2}
-\Phi^{\prime}{r^{\prime}\over r}-5{r^{\prime 2}\over r^2}
+5{r^{\prime\prime}\over r}\right ).
\end{eqnarray}

The divergences at $\epsilon\rightarrow 0$ can be eliminated by
subtracting $T_{\nu}^{\mu,{\rm div}}$, the stress-energy
corresponding to the divergent part of the effective action
derived by Christensen \cite{Christ} (simultaneously the
cosmological constant, Einstein gravity constant and coefficient
at the Weyl tensor squared
$C_{\mu\nu\lambda\rho}C^{\mu\nu\lambda\rho}$ in
the effective action are set equal to their experimental
values). Substituting our metric into the Christensen's formula
for the spinor field gives:

\begin{eqnarray}
&&\hspace{-22mm}8\pi^2r^4T_t^{t,{\rm div}}
=-8{r^4\over\epsilon^4}
+{1\over 3}{r^2\over\epsilon^2}\left (1-r^{\prime 2}+2r^{\prime
\prime}r\right )
+{1\over 60}\ln{{\Lambda\over\epsilon}}+{1\over 3}r^{\prime 2}
-{1\over 6}r^{\prime\prime}r,\\
&&\hspace{-22mm}8\pi^2r^4T_{\rho}^{\rho,{\rm div}}=+24{r^4\over
\epsilon^4}-{1\over 3}{r^2\over\epsilon^2}\left [1+r^2\left (
2\Phi^{\prime\prime}+2\Phi^{\prime 2}-2\Phi^{\prime}{r^{\prime}
\over r}-{r^{\prime 2}\over r^2}+4{r^{\prime\prime}\over r}
\right )\right ]+\nonumber\\
&&\hspace{3mm}+{1\over 60}\left (\ln{{\Lambda\over\epsilon}}-1
\right )-{1\over 36}r^2\left (\Phi^{\prime\prime}
+\Phi^{\prime 2}
+6{r^{\prime 2}\over r^2}\right ),\\
&&\hspace{-22mm}8\pi^2r^4T_{\theta}^{\theta,{\rm div}}=-8{r^4
\over\epsilon^4}+{1\over 3}{r^2\over\epsilon^2}\left (
\Phi^{\prime\prime}+\Phi^{\prime 2}-\Phi^{\prime}{r^{\prime}
\over r}+{r^{\prime\prime}\over r}\right )-{1\over 60}
\ln{{\Lambda\over\epsilon}}
-{1\over 12}r^{\prime 2}+{1\over 12}r^{\prime\prime}r.
\end{eqnarray}
The coefficient at the Weyl term is the only one which is both
logarithmically UV and (in the considered case of zero mass) IR
divergent, and $\Lambda$ is the IR cut-off. The Christensen's
procedure includes also forming half of the sum of the
components $T_{\mu\nu}$ corresponding to point separations
$\epsilon =\pm |\epsilon |$ (above formulas are given for
$\epsilon >0$), thereby only even powers of $|\epsilon |$ are
left (and also $|\epsilon|$ rather then $\epsilon$ enters the
logarithm). Subtracting $T_{\mu\nu}^{\rm div}$ from
$T_{\mu\nu}^{\rm reg}$ cancels the divergences at
$\epsilon\rightarrow 0$ (this is a useful check of our
calculation), and we finally obtain

\begin{eqnarray}
\label{T-ren}
&&\hspace{-15mm}8\pi^2r^4T_t^{t,{\rm ren}}=+{1\over 60}
\ln{{Lr\over\Lambda}}
-{1\over 72}r^{\prime 2}+{1\over 36}r^{\prime\prime}r,\\
&&\hspace{-15mm}8\pi^2r^4T_{\rho}^{\rho,{\rm ren}}
=+{1\over 60}\ln{{Lr\over\Lambda}}
+{1\over 36}\Phi^{\prime}r^{\prime}r-{1\over 72}r^{\prime 2},\\
&&\hspace{-15mm}8\pi^2r^4T_{\theta}^{\theta,{\rm ren}}=
-{1\over 60}\ln{{Lr\over\Lambda}}+{1\over 120}
+{1\over 72}r^2\left (\Phi^{\prime\prime}
+\Phi^{\prime 2}\right )
-{1\over 72}\Phi^{\prime}r^{\prime}r
+{1\over 72}r^{\prime 2}-{1\over 72}r^{\prime\prime}r.
\end{eqnarray}
Here $L\sim 1$; the IR cut off $\Lambda$ is fixed only by
experiment. The terms of higher orders in the derivatives not
taken into account in the WKB expansion here can be really
omitted if the derivatives of $\Phi$, $r$ are small as compared
to unity in the scale of $r$.

\bigskip
{\bf 3.Discussion.} The expression for the renormalised neutrino
stress-energy tensor found differs from the electromagnetic one
\cite{Kh1} only by the absolute value of the numerical
coefficients; their signs are the same. Also the difference
between the radial pressure $\tau=-T_{\rho}^{\rho,{\rm ren}}$
and the energy density $\varrho=-T_t^{t,{\rm ren}}$ at the
wormhole throat (where $r^{\prime},\Phi^{\prime}=0$ and
$r^{\prime\prime}>0$) is positive. That is, local weak energy
condition {\it at the throat} is violated. Besides that,
integrating $\tau-\varrho$ from $\rho=\rho_0<0$ to
$\rho=+\infty$ yields the sum of the two explicitly positive (in
the wormhole topology $r^{\prime}(\rho_0)<0$) terms:

\begin{equation}
\int_{\rho_0}^{\infty}{(\tau-\varrho)\exp{(-\Phi)}d\rho}=
\left.-{1\over 288\pi^2}{r^{\prime}\over r^3}\exp{(-\Phi)}
\right |_{\rho=\rho_0}+{1\over 96\pi^2}\int_{\rho_0}^{\infty}
{{r^{\prime 2}\over r^4}\exp{(-\Phi)}d\rho} > 0.
\end{equation}
That is, averaged weak energy condition \cite{MTY} is violated
as well. The difference from the electromagnetic case is in the
extent of the weak energy conditions violation which is now
$16\div 20$ times weaker.

The estimate for the wormhole size if logarithm is large
compared to unity \cite{Kh1} in the presence of $N_1$ spin 1 and
$N_{1/2}$ spin $1/2$ massless fields modifies as

\begin{equation}
r^2_0\simeq {G\over 120\pi}(4N_1+N_{1/2})
\ln{\left ({120\pi\over G}
{\Lambda^2\over 4N_1+N_{1/2}}\right )}.
\end{equation}
Typical size of the wormhole in the neutrino vacuum would be 2
times smaller than that in the electromagnetic vacuum.

It is interesting to compare our expressions for the
stress-energy of spin $1/2, 1$ massless fields with analogous
expression for the massless conformal scalar field. As given in
refs. \cite{AHS,HPS}, this expression does not contain the terms
of the second order in the derivatives of $\Phi, r$. This seems
quite natural because anomalous trace which signals that the
stress-energy is nonzero in the curved background proves to
contain no second order terms just in the conformal scalar case.
Therefore in our framework we would get $\tau-\varrho=0$ (the
leading terms with no derivatives should vanish in
$\tau-\varrho$ because we must have $\tau=\varrho$ at
$\Phi,r=const$ due to the $t\leftrightarrow\rho$ symmetry in
this case) and were to expand to the fourth order derivatives
whose values are not restricted by the wormhole topology. Thus,
violation of the weak energy conditions in the wormhole topology
with slowly varied metric for the scalar vacuum does not take
place {\em automatically} but should follow from the solution
of backreaction problem. This is the feature which distinguishes
the conformal scalar field from the nonzero spin fields.


\begin{thebibliography}{99}
\bibitem{MT}
 Morris Michael S. and Thorne Kip S., {\it Amer.J.Phys.~}{\bf
56}~(1988)~395
\bibitem{MTY}
 Morris Michael S., Thorne Kip S. and Yurtsever Ulvi, {\it
Phys.Rev.Lett.~}{\bf  61}~(1988)~1446
\bibitem{Vis}
 Visser M., {\it Lorentzian Wormholes:from Einstein to Hawking}~
(American Institute of Physics, Woodbury, 1995)
\bibitem{HPS}
 Hochberg D., Popov A. and Sushkov S.V., {\it Self-consistent
Wormhole Solutions of Semiclassical Gravity}~(Preprint LAEFF
96/25, KSPU-96-03, gr-qc/9701064) to appear in {\it
Phys.Rev.Lett.}
\bibitem{AHS}
 Anderson P.A., Hiscock W.A. and Samuel D.A., {\it Phys.Rev.~}
{\bf D51}~(1995)~4337
\bibitem{AHT}
 Anderson P.A., Hiscock W.A. and Taylor B.E., {\it Stress-energy
of a quantized scalar field in static wormhole spacetimes}
(Preprint gr-qc/9608038, August 1996), submitted to
{\it Phys.Rev.~}{\bf D}
\bibitem{Kh}
 Khatsymovsky V.M., {\it Phys.Lett.~}{\bf B320}~(1994)~234
\bibitem{Christ}
 Christensen S.M., {\it Phys.Rev.~}{\bf D17}~(1978)~946
\bibitem{FW}
 Flanagan E.E. and Wald R.M., {\it Phys.Rev.~}{\bf D54}~(1996)
~6233
\bibitem{Kh1}
 Khatsymovsky V.M., {\it Towards the possibility of a
self-maintained vacuum-traversible wormhole}, to appear in {\it
Phys.Lett.~}{\bf B}
\bibitem{FR}
 Ford L.H. and Roman T.A., {\it Phys.Rev.~}{\bf D53}~(1996)~5496
\bibitem{DeWitt}
 DeWitt B.S., {\it Phys.Rep.~}{\bf 19C}~(1975)~295
\bibitem{DeWitt1}
 DeWitt B.S., {\it The Dynamical Theory of Groups and Fields}
(Gordon and Breach, New York, 1965)
\end{thebibliography}
\end{document}